\documentclass[journal]{IEEEtran}
\usepackage{amsmath,amssymb}
\usepackage[T1]{fontenc}
\usepackage{graphicx}
\usepackage{array,booktabs}
\usepackage{cite}
\usepackage{url}
\usepackage[hidelinks]{hyperref}

\providecommand{\ul}[1]{#1}
\renewcommand{\thetable}{\arabic{table}}
\usepackage{newunicodechar}
\newunicodechar{×}{\ensuremath{\times}}
\newunicodechar{−}{\ensuremath{-}}
\begin{document}
\hypersetup{pdftitle={Resource Adequacy with Renewables and Storage—Part II: Numerical Validation and the Capacity Accreditation Gap},pdfauthor={Philip Q. Hanser},pdfkeywords={resource adequacy, energy storage, capacity accreditation}}
\title{Resource Adequacy with Renewables and Storage---Part II: Numerical Validation and the Capacity Accreditation Gap}
\author{Philip~Q.~Hanser,~\IEEEmembership{Life~Member,~IEEE}\thanks{P.~Q. Hanser is a Principal Emeritus of The Brattle Group, Boston, MA 02108 USA (e-mail: Philip.Hanser@affiliate.brattle.com). The views expressed are the author's own and do not represent the positions of The Brattle Group or of any client. This work has been submitted to the IEEE for possible publication. Copyright may be transferred without notice, after which this version may no longer be accessible.}}
\maketitle
\begin{abstract}
This paper tests the decomposition set out in Part I of this study on 4 years of hourly Spanish system data and derives its consequence for capacity accreditation. Under high renewable penetration and energy-limited storage, expected energy not served and loss-of-load hours for a single reservoir split exactly into two terms: a power-limited term, which the classical capacity-outage convolution computes, and an energy-limited term, the expected cumulative energy that the empty reservoir fails to deliver. A discrete-time solver for the reflected process reproduces the closed-form benchmark of Part I and computes the split on the recorded data. With a 4-hour reservoir, the energy-limited term accounts for 98\% of expected energy not served, and it vanishes only when the reservoir holds enough energy to cover the cumulative drawdown of the multi-day droughts. Effective load-carrying capability holds fixed a count of days. That count registers how often the energy-limited term produces a shortfall but not how deep the shortfall runs, so a credit matched on it misstates the unserved energy the term carries. On these data, a 4-hour battery meets the one-day-in-ten-years standard while carrying nearly 10 times the unserved energy that its accredited model reports. The error follows the energy-limited term's share of expected energy not served and vanishes with it.
\end{abstract}
\emph{This paper is Part II of a four-part study. Part I {[}1{]} develops the theory and is cited throughout in the form {[}1, Sec. 8.3{]}; unqualified section numbers refer to this paper.}

\section*{1. Introduction}

Resource-adequacy assessment asks how much firm demand a system can serve at a target reliability. Part I {[}1{]} shows that under high renewable penetration and energy-limited storage the additive loss-of-load and energy-not-served metrics for a single reservoir decompose exactly into two terms: a memoryless, power-limited term equal to the classical capacity-outage convolution, with the load axis shifted by the storage power rating, and an energy-limited term equal to the expected regulator of the state of charge at its empty boundary. The power-limited term is the fragment the profession has computed for six decades; the energy-limited term is what storage adds, and in a system whose binding events are multi-day renewable droughts it is the term that binds.

This paper puts that identity on real data and follows it into the reporting layer. It runs the reflected-process solver on 4 years of hourly Spanish system data, validates it against the closed-form benchmark of {[}1, Appendix A{]}, exhibits the mechanism on an actual December 2016 drought, and reports the exact split of both additive metrics (Section 2). It then turns the decomposition on capacity accreditation and shows that effective load-carrying capability (ELCC), which holds a count of days fixed, cannot register the depth of the term that binds: on these data a 4-hour battery meets the one-day-in-ten-years standard while carrying nearly 10 times the unserved energy its accredited model reports (Section 3). Section 4 concludes. Replication code and data accompany the paper.

\section*{2. Numerical illustration on high-penetration data}

Part I states an identity and two consequences. This section tests them on a real, weather-driven record: the solver reproduces the closed-form benchmark of {[}1, Appendix A{]}, the decomposition behaves as the theory requires, and the mechanism generating the boundary term is the one {[}1, Sec. 7{]} describes. The accreditation consequence is deferred to Section 3, where the numbers accompany the argument they support.

\subsection*{2.1 Data and scenario}

The test record is 4 years of hourly Spanish system data, load and onshore wind and solar generation, published by the European Network of Transmission System Operators for Electricity (ENTSO-E) for 2015 through 2018, or 35,023 complete hours. Spain is useful because the record is real: it carries the wind, solar, and load correlation, the seasonal non-stationarity, and the multi-day low-renewable spells that a fitted model would have to reproduce.

Renewable capacity is scaled by a common factor to roughly 90\% of annual load energy while the hourly shapes are held fixed, so the joint tails and the droughts survive the scaling untouched. A thermal fleet of twenty-six units of 1,000 MW is added, with a 6\% forced-outage rate and a 48-hour mean time to repair, so outages persist along the path. A single reservoir of 12 GW is dispatched under the reliability-greedy rule of {[}1, Sec. 7.5{]}, with duration swept from 2 to 256 hours. Mean load is 28.7 GW and the peak 41.0 GW. Metrics are Monte Carlo averages over 200 independent thermal-outage histories drawn against the one weather record. Loss-of-load expectation (LOLE) is the expected number of days per year containing at least one shortfall hour, the definition adequacy standards use; unlike expected energy not served (EENS) and loss-of-load hours (LOLH), it is not additive and does not split into the two terms of {[}1, Sec. 8.3{]}. Section 2.6 states the limits of the exercise.

The capacity credits this scenario produces, reported in Table 3, invite comparison with the accreditation values system operators publish, and the comparison needs care. PJM's ELCC class ratings for the 2026/2027 base residual auction are 50\% of installed capacity for 4-hour storage, rising to 58\%, 62\%, and 72\% at 6, 8, and 10 hours {[}2{]}. Table 3 gives 73\% at 4 hours and 97\% at 8. The levels are not comparable, and should not be read as agreement or as disagreement: a class rating is the marginal contribution of an increment of storage to a system already carrying a large fleet and dominated by winter risk, while Table 3 reports the equivalent firm capacity of a single reservoir in a system whose binding events are evening ramps.

What is comparable is the shape of the curve and the reason given for it. Credit rises with duration in both, and PJM attributes the decline in its ratings to winter risk periods that ``tend to last multiple hours,'' and to storage being unable to recharge during those events {[}3{]}. That is the criterion of {[}1, Sec. 8.7{]}---stored energy measured against the energy the binding event draws down---arrived at independently, on a different system, and visible in a market that has cleared.

\subsection*{2.2 Solver validation}

Part I supplies a general transient system {[}1, Sec. 8.4{]} and a closed form for the two-regime case {[}1, Appendix A{]}, the second existing so the first can be checked. Fig. 1 performs the check. The discrete reflected recursion of {[}1, Sec. 8.5{]}, run at a time step of 0.002 in the benchmark's time units, reproduces the closed-form empty-boundary probability across reservoir sizes to within about 2\% and converges to it as the time step shrinks. The same solver produces everything that follows.

\begin{figure*}[!t]
\centering
\includegraphics[width=\textwidth]{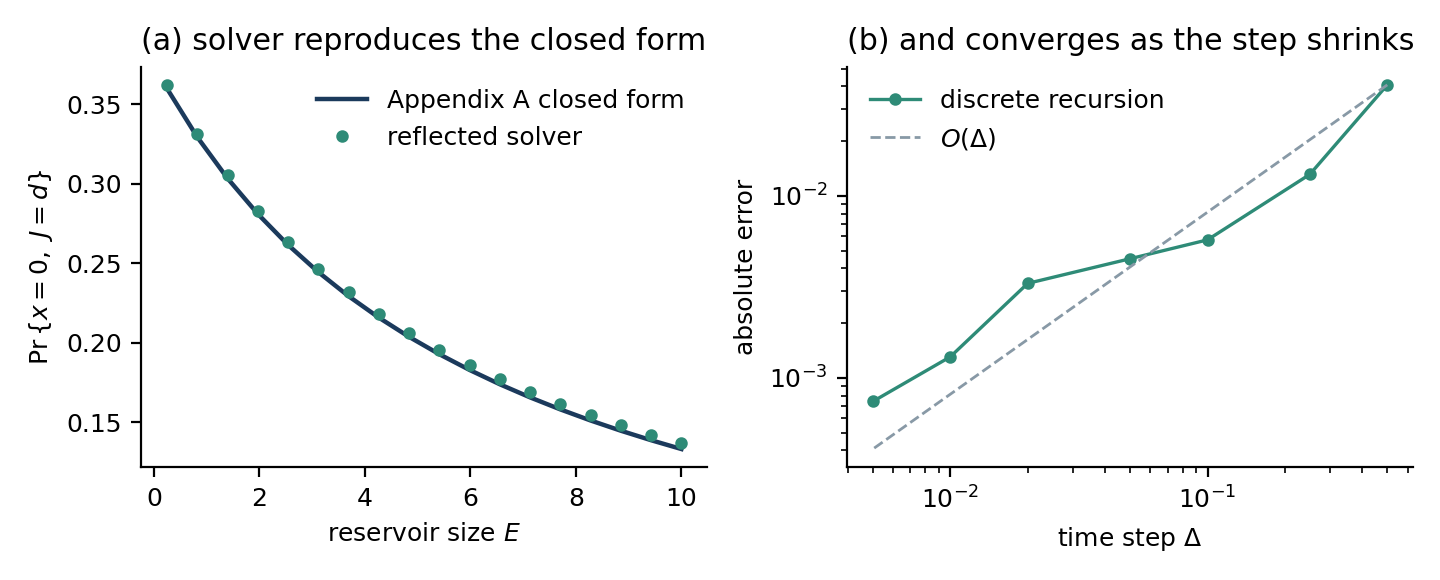}
\caption{\emph{The reflected-process solver reproduces the closed form of {[}1, Appendix A{]} (left) and converges under refinement of the time step (right).}}
\end{figure*}
\subsection*{2.3 What a renewable drought does to the reservoir}

The mechanism of {[}1, Sec. 7{]} is easiest to see in a single event. Fig. 2 shows the worst multi-day window in the record, a December 2016 winter drought, against a 2-hour reservoir. Firm plus renewable supply falls short of load for several days, and the deficit sits almost entirely within the 12-GW power rating, so it is an energy problem and not a power problem. The classical calculation, which credits storage at its full power rating in every hour, records none of it.

The reservoir is not pinned at empty throughout the window, and that detail governs how the metrics read. It drains to empty on three successive evenings, each depletion ended by the next day's residual solar partially refilling it and reopened by the next evening's ramp. The drought prevents full recovery; the diurnal cycle divides the shortfall into nightly pieces. Every hour at empty in deficit adds to the regulator of {[}1, Sec. 7.4{]}, and across this window the expected energy-limited total is 23.4 GWh, none of it power-limited.

\begin{figure}[!t]
\centering
\includegraphics[width=\columnwidth]{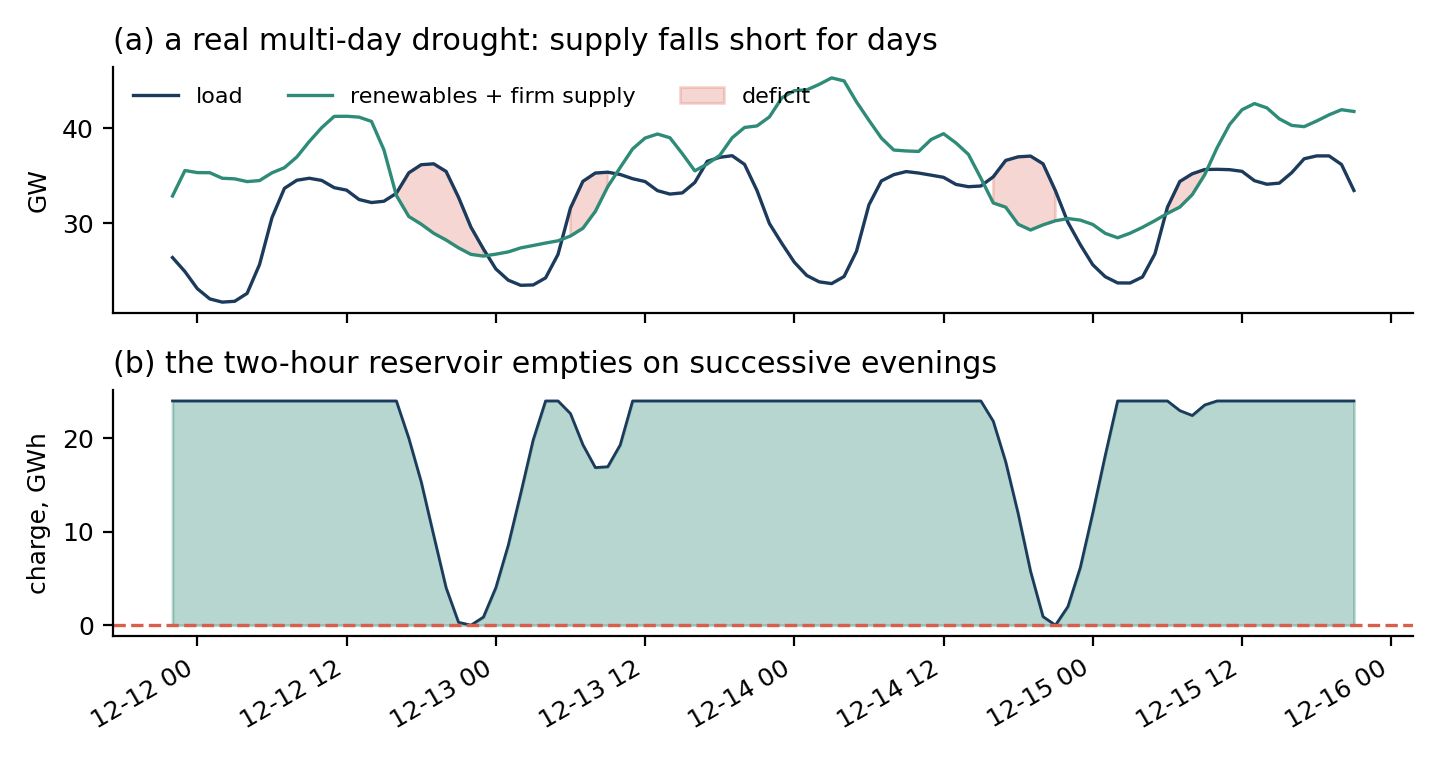}
\caption{\emph{A real multi-day drought empties a 2-hour reservoir on successive evenings, December 11 to 15, 2016. The shortfall is energy-limited, 23.4 GWh across the window in expectation with none of it power-limited, and invisible to a static convolution. A reservoir of 8 hours or more rides through the same window without reaching empty.}}
\end{figure}
\subsection*{2.4 The decomposition on real data}

Table 1 sets the reflected process against the crudest representation of storage: crediting it at full power in every hour, so only a deficit exceeding 12 GW can produce a shortfall. This is a bound and not a practice, and the accreditation apparatus of Section 3 exists in order not to use it. The contrast is reported because it isolates the boundary term, everything above the flat classical row being the energy-limited term, and because the ordering across the three metrics survives any baseline, including the reasonable one of Section 3.

\begin{table*}[!t]
\caption{Storage credited as firm power in every hour, a bound and not a practice, against reflected-process adequacy metrics. Understatement factors in parentheses. Expected energy not served (EENS) in GWh/yr; loss-of-load hours (LOLH) in h/yr; loss-of-load expectation (LOLE) in d/yr. Durations of 128 and 256 hours are identical to 64 hours and are omitted.}
\centering\footnotesize
\begin{tabular}{@{} >{\raggedright\arraybackslash}p{\dimexpr 0.1324\linewidth-2\tabcolsep\relax} >{\raggedright\arraybackslash}p{\dimexpr 0.1446\linewidth-2\tabcolsep\relax} >{\raggedright\arraybackslash}p{\dimexpr 0.1446\linewidth-2\tabcolsep\relax} >{\raggedright\arraybackslash}p{\dimexpr 0.1446\linewidth-2\tabcolsep\relax} >{\raggedright\arraybackslash}p{\dimexpr 0.1446\linewidth-2\tabcolsep\relax} >{\raggedright\arraybackslash}p{\dimexpr 0.1446\linewidth-2\tabcolsep\relax} >{\raggedright\arraybackslash}p{\dimexpr 0.1446\linewidth-2\tabcolsep\relax}@{}}
\toprule
Storage duration & EENS classical & EENS reflected & LOLH classical & LOLH reflected & LOLE classical & LOLE reflected \\
\midrule
\bottomrule
2 h & 0.21 & \textbf{55.71} (270×) & 0.21 & \textbf{14.77} (71×) & 0.16 & \textbf{4.25} (27×) \\
4 h & 0.21 & \textbf{12.99} (63×) & 0.21 & \textbf{2.94} (14×) & 0.16 & \textbf{0.94} (6.0×) \\
8 h & 0.21 & \textbf{0.87} (4.2×) & 0.21 & \textbf{0.32} (1.6×) & 0.16 & \textbf{0.19} (1.2×) \\
16 h & 0.21 & 0.21 & 0.21 & 0.21 & 0.16 & 0.16 \\
32 h & 0.21 & 0.21 & 0.21 & 0.21 & 0.16 & 0.16 \\
64 h & 0.21 & 0.21 & 0.21 & 0.21 & 0.16 & 0.16 \\
\end{tabular}
\end{table*}
The firm-power values are flat and small in every column, a fifth of a gigawatt-hour per year and a fifth of an hour, because the deficit rarely exceeds the 12-GW rating. The reflected values are far larger at the durations that matter. At 4 hours, the canonical lithium-ion duration, the contrast understates EENS 63-fold, LOLH 14-fold, and LOLE 6-fold (Fig. 3). The columns converge only at 16 hours and beyond, where the reservoir rides through the droughts and the energy-limited term vanishes identically on this record. That is the criterion of {[}1, Sec. 8.7{]} satisfied on real data.

\begin{figure*}[!t]
\centering
\includegraphics[width=\textwidth]{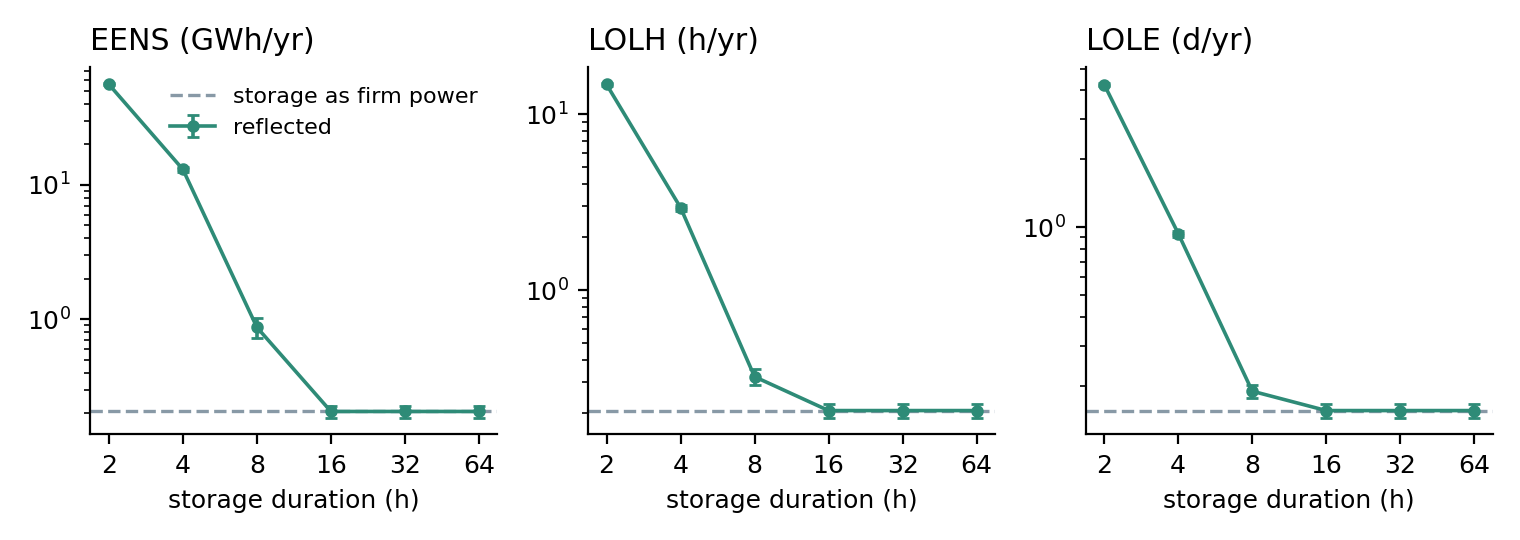}
\caption{\emph{The firm-power contrast understates all three metrics at realistic durations, and the coarser the metric the smaller the apparent error. Bars show Monte Carlo standard errors across the 200 thermal-outage histories.}}
\end{figure*}
EENS is the product of how often shortfalls begin (loss-of-load events), how long each lasts, and how deep it runs: events per year, times hours per event, times megawatts unserved per shortfall hour. The metrics of Table 1 are this product truncated from the right. LOLH keeps frequency and duration and discards depth; LOLE keeps frequency and, through the day count, a residue of duration. At the 2-hour reservoir the classical representation is wrong in all three factors: the shortfall it reports is 28.4 times too rare, 2.5 times too short, and 3.8 times too shallow. The errors multiply to 270, the expected-energy ratio; drop the depth factor and they multiply to 71, the loss-of-load-hours ratio. Each coarsening of the metric discards a factor in which the classical view is wrong, and so conceals part of its own error. LOLE, the metric adequacy standards are written in, discards the most and reveals the least.

Fig. 4 splits both additive metrics into their exact components and is the direct test of {[}1, Sec. 8.3{]}. The power-limited term is the smaller contribution at every short duration, in hours as in megawatt-hours. The empty-boundary term supplies nearly all of the risk until the reservoir is large enough for depletion to cease: at 4 hours it is 98\% of EENS and 93\% of LOLH. The convolution survives in both metrics, but as the smaller fragment of a larger object whose binding term is a boundary functional, a regulator where the shortfall is weighted by energy and an occupation time where it is merely counted.

\begin{figure*}[!t]
\centering
\includegraphics[width=\textwidth]{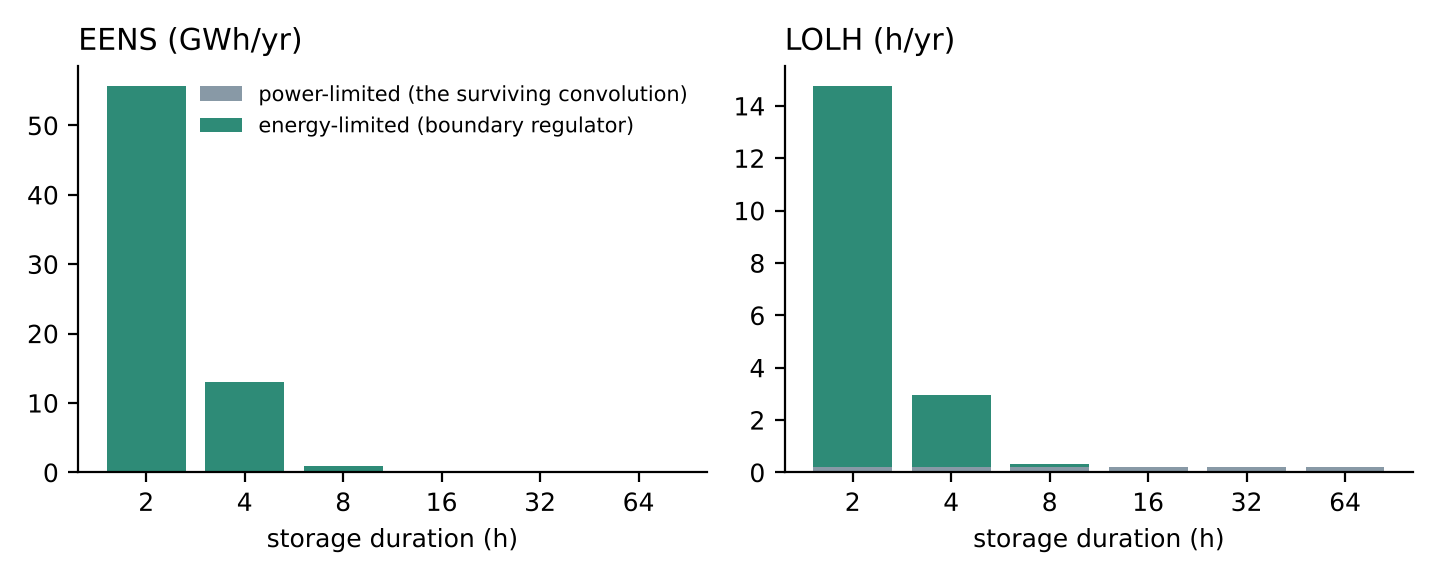}
\caption{\emph{EENS and LOLH each split exactly into a memoryless convolution plus an empty-boundary term, the first and zeroth moments of the same boundary measure. The boundary term dominates at short duration.}}
\end{figure*}
Table 2 supplies the frequency and duration factors and locates the energy-limited term more precisely than the drought narrative alone. The classical convolution produces 0.16 events per year, none longer than 4 hours: ordinary coincidences of thermal outage and peak load, and all it can produce. The reflected process at 2 hours produces 4.41 events per year, averaging 3.4 hours, the longest running 17 hours (Fig. 5). The events are therefore not multi-day outages. Events per year and shortfall days per year track each other almost exactly, 4.41 against 4.25, which pins the structure down. The energy-limited term is not one sustained failure spanning a drought but a run of single evening failures on consecutive days, each ended by the next morning's partial recharge. A metric that counts days sees the drought; a metric that counts events sees the nights; only the energy-weighted boundary functional sees both.

\begin{table*}[!t]
\caption{Loss-of-load events and their duration. Event statistics are pooled over the 200 thermal-outage histories. Durations of 128 and 256 hours are identical to 64 hours and are omitted.}
\centering\footnotesize
\begin{tabular}{@{} >{\raggedright\arraybackslash}p{\dimexpr 0.1324\linewidth-2\tabcolsep\relax} >{\raggedright\arraybackslash}p{\dimexpr 0.1446\linewidth-2\tabcolsep\relax} >{\raggedright\arraybackslash}p{\dimexpr 0.1446\linewidth-2\tabcolsep\relax} >{\raggedright\arraybackslash}p{\dimexpr 0.1446\linewidth-2\tabcolsep\relax} >{\raggedright\arraybackslash}p{\dimexpr 0.1446\linewidth-2\tabcolsep\relax} >{\raggedright\arraybackslash}p{\dimexpr 0.1446\linewidth-2\tabcolsep\relax} >{\raggedright\arraybackslash}p{\dimexpr 0.1446\linewidth-2\tabcolsep\relax}@{}}
\toprule
Storage duration & Events/yr classical & Events/yr reflected & Mean event (h) classical & Mean event (h) reflected & Longest (h) classical & Longest (h) reflected \\
\midrule
\bottomrule
2 h & 0.16 & \textbf{4.41} & 1.33 & \textbf{3.35} & 4 & \textbf{17} \\
4 h & 0.16 & \textbf{0.93} & 1.33 & \textbf{3.15} & 4 & \textbf{12} \\
8 h & 0.16 & \textbf{0.20} & 1.33 & \textbf{1.64} & 4 & \textbf{6} \\
16 h & 0.16 & 0.16 & 1.33 & 1.33 & 4 & 4 \\
32 h & 0.16 & 0.16 & 1.33 & 1.33 & 4 & 4 \\
64 h & 0.16 & 0.16 & 1.33 & 1.33 & 4 & 4 \\
\end{tabular}
\end{table*}
\begin{figure*}[!t]
\centering
\includegraphics[width=\textwidth]{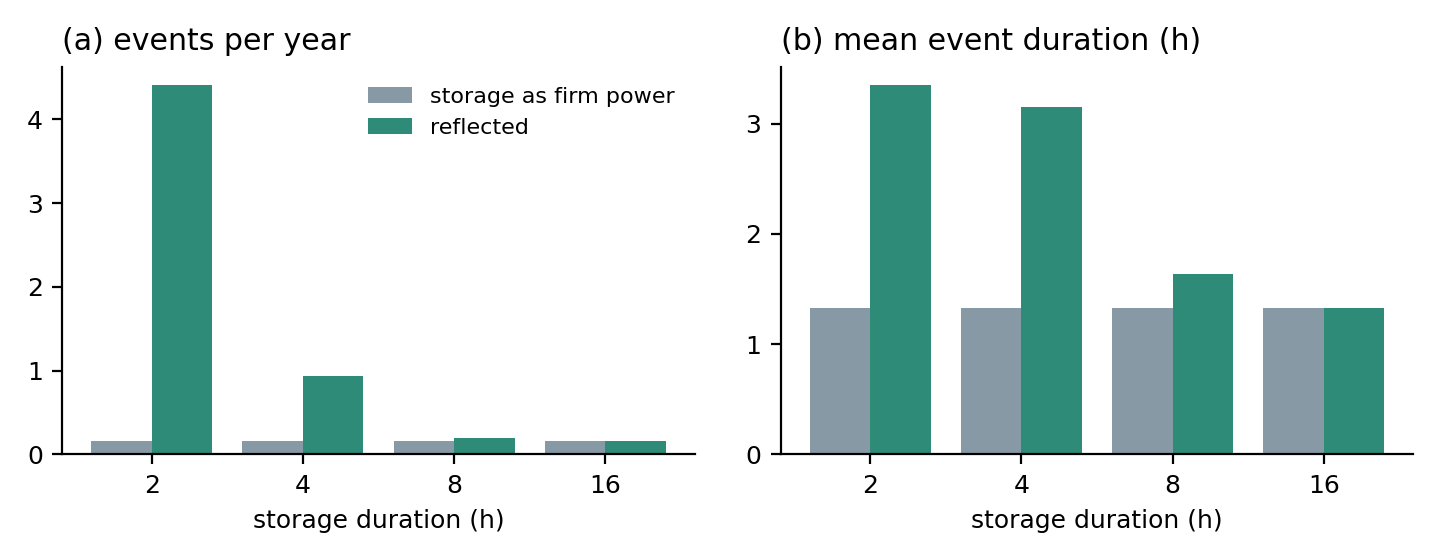}
\caption{\emph{The energy-limited term is made of more outages and longer outages, but not of one long outage. (a) The classical convolution produces nothing beyond 4 hours; the reflected process produces evening events out to 17 hours. (b) Event count and mean event length converge on the classical values once the reservoir reaches 16 hours.}}
\end{figure*}
\subsection*{2.5 The screening test exercised}

Part I proposes the decomposition as a screen {[}1, Sec. 8.6{]}: compute the power-limited term by quadrature, compare the reservoir's energy rating against the largest cumulative energy drawdown the deficit path imposes {[}1, Sec. 8.7{]}, and decide whether a sequential simulation is required. On the Spanish record the screen gives the right answer, and the choice of statistic matters: drought length alone would have given the wrong one.

The power-limited term is 0.21 GWh/yr at every duration, computed by quadrature in negligible time. The binding low-renewable episodes in the Spanish record run to several days, the deepest window of Fig. 2 spanning the better part of a week, so a comparison of duration against drought length would screen in every reservoir shorter than several days, the 16-hour reservoir included. The drawdown does not. With thermal capacity held at its expected availability, one pass over the record puts the largest drawdown at 76 GWh, 6.3 hours of the 12-GW rating; across the 200 outage histories it ranges from 49 to 143 GWh, or 4.1 to 11.9 hours. Within a drought, daily solar recharge and deficits shallower than the power rating hold the drawdown to a fraction of what the episode's length suggests. The screen therefore places the 2- and 4-hour reservoirs in, since every history overdraws them; marks the 8-hour reservoir as marginal, overdrawn in 22\% of histories; and screens the 16-hour and longer reservoirs out, since no history draws down more than 143 GWh against the 192 GWh a 16-hour reservoir holds. All three calls are correct, and the last exactly so: at 16 hours and beyond, Table 1 shows the reflected and classical columns agreeing to the reported precision in all three metrics, because the boundary term is zero on every path. A planner running the screen would have known, before committing to any simulation, that the 4-hour case required one and the 32-hour case did not.

\subsection*{2.6 Scope and caveats}

The illustration establishes a point of method. On a real record with weather-driven joint structure, the identity of {[}1, Sec. 8.3{]} holds numerically against a validated solver, the boundary term supplies the great majority of the risk at the durations now being built, and the criterion of {[}1, Sec. 8.7{]} correctly predicts where it does not.

Its limits are equally specific. The reported standard errors reflect sampling over thermal-outage histories only. The weather record is a single 4-year realization, so the sampling error in the drought frequency driving the energy-limited term is not quantified and is likely the larger of the two. Renewable capacity is scaled to a counterfactual penetration, and the thermal fleet, forced-outage rates, and storage sizing are stylized. The dispatch is the reliability-greedy rule, which for a single reservoir serves at least as much energy as any other policy on every path {[}1, Sec. 8.5{]}, so the reflected values are a lower bound on the path-dependent risk under economic dispatch. None of this affects the identity, which holds exactly and path by path. It affects the magnitudes, which illustrate what is at stake rather than estimating it for any system.

\section*{3. Accreditation under a boundary functional}

Applied to the reporting layer, the decomposition shows what capacity accreditation cannot register. ELCC is defined as the increment of firm demand a resource lets the system carry while holding LOLE fixed. LOLE is a count of days containing at least one shortfall hour: a functional of daily maxima. The energy-limited term of {[}1, Sec. 8{]} is an additive functional of the path, the empty-boundary regulator. A count of daily maxima registers whether that functional is positive on a given day, and Table 1 shows LOLE rising 6-fold with it at 4 hours. It does not register how large the functional is, because a day holding a 1-hour shortfall and one holding an 8-hour shortfall at 4 times the depth enter the count identically. Matching LOLE therefore does not match EENS, and not for want of resolution: the quantity accreditation holds fixed carries no information about depth.

The credit computed below is the equivalent firm capacity: the firm capacity whose addition reproduces the reservoir's LOLE. Garver's ELCC, the increment of load served at fixed LOLE, is a distinct metric that generally yields a different number {[}4{]}. Both hold the same count fixed, so the argument applies to either; the tables report the first because it is the quantity a capacity credit assigns.

The consequence is not that accreditation errs at random but that it errs in a direction, by an amount the decomposition predicts. Credit a reservoir at the firm capacity that reproduces its LOLE---which is what a capacity credit is---and the accredited system agrees on LOLE by construction and disagrees on EENS. Table 3 and Fig. 6 report the size on the data of Section 2. For each duration the reservoir's LOLE is computed on the reflected process, the firm capacity reproducing that expectation is found, and the accredited system's unserved energy is read off. The capacity credits---50\%, 73\%, 97\%, and 100\% of nameplate at 2, 4, 8, and 16 hours---trace the familiar duration-saturation curve, which is a check on the scenario rather than a result. The result is the last column: LOLE agrees to within 0.01 day per year by construction, and EENS does not agree at all.\footnote{The scenario is the one specified in Section 2.1 of this paper. The accreditation errors quoted here use the conservative credit---the largest credit whose accredited system is no more reliable than the truth---so they are lower bounds.} The pattern is not a coincidence of the scenario. The boundary term supplies 100\%, 98\%, 76\%, and 0\% of unserved energy at those same durations, and the accreditation error tracks that share and vanishes with it. Where the reservoir is long enough to ride through the droughts, the energy-limited term is identically zero, the system is memoryless, and the capacity credit is exactly right. Accreditation fails to the extent that the term whose depth it cannot see is the term that binds.

\begin{table*}[!t]
\caption{Accreditation error at the operating point. Storage is credited at the firm capacity reproducing its LOLE, and the accredited model is then read. LOLE in d/yr, EENS in GWh/yr. Mean unserved power per shortfall hour---the depth the count cannot see---is 3.77, 4.41, 2.71, and 1.00 GW on the reflected process against 1.56, 1.45, 0.98, and 0.99 GW on the accredited model.}
\centering\footnotesize
\begin{tabular}{@{} >{\raggedright\arraybackslash}p{\dimexpr 0.1310\linewidth-2\tabcolsep\relax} >{\raggedright\arraybackslash}p{\dimexpr 0.1310\linewidth-2\tabcolsep\relax} >{\raggedright\arraybackslash}p{\dimexpr 0.1310\linewidth-2\tabcolsep\relax} >{\raggedright\arraybackslash}p{\dimexpr 0.1548\linewidth-2\tabcolsep\relax} >{\raggedright\arraybackslash}p{\dimexpr 0.1310\linewidth-2\tabcolsep\relax} >{\raggedright\arraybackslash}p{\dimexpr 0.1548\linewidth-2\tabcolsep\relax} >{\raggedright\arraybackslash}p{\dimexpr 0.1666\linewidth-2\tabcolsep\relax}@{}}
\toprule
Storage duration & Capacity credit & LOLE true & LOLE accredited & EENS true & EENS accredited & Understated \\
\midrule
\bottomrule
2 h & 6.06 GW (50\%) & 4.249 & 4.256 & 55.71 & 17.85 & 3.1× \\
4 h & 8.75 GW (73\%) & 0.936 & 0.941 & 12.99 & 2.97 & 4.4× \\
8 h & 11.66 GW (97\%) & 0.189 & 0.189 & 0.87 & 0.28 & 3.1× \\
16 h & 12.01 GW (100\%) & 0.155 & 0.155 & 0.21 & 0.20 & 1.0× \\
\end{tabular}
\end{table*}
The mechanism is depth, and it explains why no refinement of the count repairs the matter. A reservoir fails deep: it fails when it is empty, and an empty reservoir surrenders the whole deficit, so the shortfall runs at the full depth of the drought evening that caused it. Firm capacity fails shallow: it fails only when the deficit exceeds it, so its shortfalls are marginal by construction, about 1.0 to 1.6 gigawatts per shortfall hour beyond a threshold. Matched on the day count, the two systems differ in depth by a factor of 2.4 to 3.0---and depth is the one factor a count cannot carry. Table 4 confirms that the diagnosis is structural rather than a quarrel with day-counting: matching LOLH instead of days---a finer metric on the same axis---still understates unserved energy by 2.4 to 2.9 times, because hours cannot see depth either. Unserved energy factors into event frequency, event duration, and shortfall depth; each coarsening of the metric discards a factor from the right, and no metric on the left of depth can be calibrated to recover it.

\begin{table*}[!t]
\caption{Control. Matching LOLH instead of days---a finer metric on the same axis---still understates unserved energy. The defect is not day-counting: no count-based metric sees depth.}
\centering\footnotesize
\begin{tabular}{@{} >{\raggedright\arraybackslash}p{\dimexpr 0.2000\linewidth-2\tabcolsep\relax} >{\raggedright\arraybackslash}p{\dimexpr 0.2000\linewidth-2\tabcolsep\relax} >{\raggedright\arraybackslash}p{\dimexpr 0.2000\linewidth-2\tabcolsep\relax} >{\raggedright\arraybackslash}p{\dimexpr 0.2000\linewidth-2\tabcolsep\relax} >{\raggedright\arraybackslash}p{\dimexpr 0.2000\linewidth-2\tabcolsep\relax}@{}}
\toprule
Storage duration & LOLE-matched credit & EENS understated & LOLH-matched credit & EENS understated \\
\midrule
\bottomrule
2 h & 6.06 GW & 3.1× & 5.65 GW & 2.4× \\
4 h & 8.75 GW & 4.4× & 8.16 GW & 2.9× \\
8 h & 11.66 GW & 3.1× & 11.56 GW & 2.8× \\
\end{tabular}
\end{table*}
Nearer the standard the error grows rather than subsides. Table 5 adds firm backup until each storage system meets the one-day-in-ten-years criterion and accredits it on the LOLE it then achieves. A 4-hour battery earns 83\% of nameplate, meets the standard, and carries 9.7 times the unserved energy its accredited model reports. The direction is intelligible: at the margin the accredited system's shortfalls become more nearly marginal, while the reservoir's remain drought-deep, so the depth ratio widens as the criterion tightens. A planner is therefore least well served by the instrument where the instrument is being read.

\begin{table*}[!t]
\caption{At the one-day-in-ten-years standard. Firm backup is added until each storage system meets the standard, and the storage is then accredited on the LOLE it achieves. Because that metric is a step function of firm capacity near the standard, the credits are conservative---the largest whose accredited system is no more reliable than the truth, an accredited 0.144 against a true 0.083 to 0.098---so the ratios are lower bounds.}
\centering\footnotesize
\begin{tabular}{@{} >{\raggedright\arraybackslash}p{\dimexpr 0.1342\linewidth-2\tabcolsep\relax} >{\raggedright\arraybackslash}p{\dimexpr 0.1342\linewidth-2\tabcolsep\relax} >{\raggedright\arraybackslash}p{\dimexpr 0.1342\linewidth-2\tabcolsep\relax} >{\raggedright\arraybackslash}p{\dimexpr 0.1342\linewidth-2\tabcolsep\relax} >{\raggedright\arraybackslash}p{\dimexpr 0.1342\linewidth-2\tabcolsep\relax} >{\raggedright\arraybackslash}p{\dimexpr 0.1585\linewidth-2\tabcolsep\relax} >{\raggedright\arraybackslash}p{\dimexpr 0.1708\linewidth-2\tabcolsep\relax}@{}}
\toprule
Storage duration & Firm backup & LOLE achieved & Capacity credit & EENS true & EENS accredited & Understated \\
\midrule
\bottomrule
2 h & 4.91 GW & 0.095 & 7.65 GW (64\%) & 0.699 & 0.105 & 6.7× \\
4 h & 2.54 GW & 0.098 & 10.02 GW (83\%) & 1.020 & 0.105 & 9.7× \\
8 h & 0.56 GW & 0.083 & 12.00 GW (100\%) & 0.371 & 0.105 & 3.5× \\
\end{tabular}
\end{table*}
\begin{figure*}[!t]
\centering
\includegraphics[width=\textwidth]{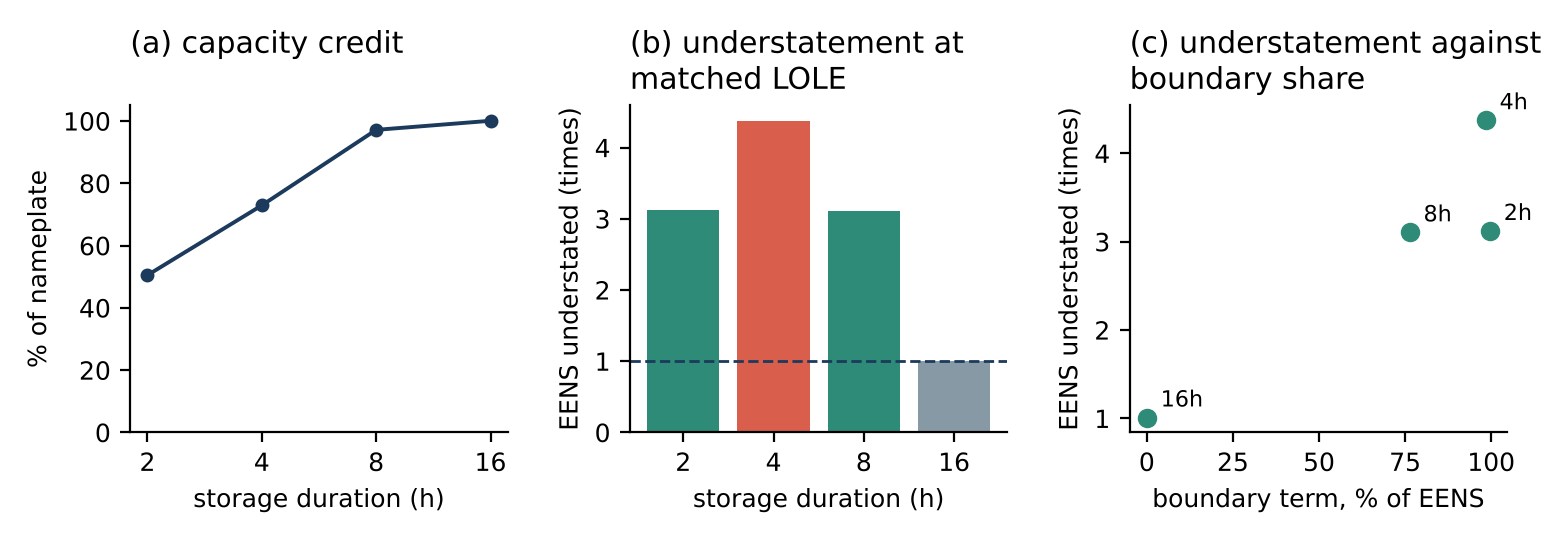}
\caption{\emph{Accreditation under a boundary functional. (a) The capacity credit rises with duration, tracing the familiar saturation curve. (b) At matched LOLE, EENS is understated 3.1, 4.4, and 3.1 times at 2, 4, and 8 hours, and not at all at 16. (c) The understatement arises only when the boundary term does: at 2 to 8 hours, where the term supplies 76\% to 100\% of unserved energy, EENS is understated 3.1 to 4.4 times; at 16 hours the term vanishes and accreditation is exact.}}
\end{figure*}
One further property of the metric concerns the observation of {[}1, Sec. 6{]} that adequacy is dominated by a handful of weather years. Near the standard, LOLE is not merely coarse but granular: it is a step function of firm capacity. On the data of Section 2, 200 megawatts of credit nearly halve it, from 0.144 to 0.075 days per year, a step of about two-thirds of the standard itself. The cause is the one {[}1, Sec. 5{]} identified as a virtue in the estimation context: the weather record is shared across every outage history, so the worst drought day is a shortfall day in all of them at once and leaves all of them at the same capacity. The quantity practice conditions on correctly is the quantity that makes the reported metric jump. At one day in ten years the standard is being read off a step function whose steps are on the order of the standard itself.

One consequence carries into any reporting built on the above. Under the dependence at issue, ELCC ceases to be additive: the capability of a wind-plus-solar portfolio is not the sum of the individual capabilities, owing to the diversity interaction, so marginal, last-in, and average allocations diverge. Capacity accreditation is therefore a consequence of the analysis, not its starting point: it requires the retained multivariate object and, once storage enters, the path-dependent term. The IEEE task force treatment of wind capacity value works this through {[}4{]}. Non-additivity and depth-blindness are the same defect met twice: a scalar credit is asked to summarize an object---a multivariate law in the first case, a boundary functional in the second---that no scalar determines.

\subsection*{3.1 Where this leaves ELCC}

System operators accredit storage with ELCC, including in systems already in the regime this paper concerns. The results above weigh against it, in three claims of decreasing generality that fail at different rates.

\emph{The depth claim is definitional and admits no implementation.} ELCC holds a count fixed; the binding term is an additive functional of the path; a count registers whether that functional is positive but not how large it is. Nothing about how the count is computed, over how many weather years or with what outage model, changes this. Table 4 makes the point in the only way available, by refining the count and showing the error persists. The claim cannot be answered by improving an implementation.

\emph{The conditioning claim is quantitative and worsens as the system decarbonizes.} ELCC reports a megawatt quantity, which presumes megawatts are the scarce thing. Where the binding event is a multi-day energy deficit they are not, and the metric becomes ill-conditioned in a precise sense: what it reports moves little in response to large changes in the property determining reliability. Table 3 shows a 4-hour and a 16-hour reservoir of identical nameplate accredited at 73\% and 100\%, while their unserved energy differs by a factor of 60. The credit is not merely imprecise; it is responding to the wrong variable. Part I supplies the reason {[}1, Sec. 8.7{]}: the boundary term is governed by stored energy measured against the drawdown of the droughts, and a megawatt figure carries no information about either.

\emph{The policy claim concerns what the metric cannot represent at all.} Part I identifies anticipatory dispatch {[}1, Sec. 7.5{]}, pre-charging storage against a days-ahead dunkelflaute forecast, as what separates a reservoir dispatched for cost from one held for reliability, and as the policy that sets how much energy the reservoir carries into the binding event. A capacity credit has nowhere to put this. It is a single number attached to a resource, while forecast skill and lead time belong to the operator's information; two identical batteries in systems with different forecast horizons contribute differently to reliability, and no megawatt figure distinguishes them. The same holds, less dramatically, of the reliability-greedy rule, which is already an assumption about behavior rather than a property of hardware. Stephen et al.~{[}5{]} document how far capacity credit moves under dispatch assumptions. The point here is stronger than sensitivity: the object being credited is a policy as much as a resource, and the credit has no argument for it.

Matching unserved energy instead of days does not repair the credit; it moves the error. Tables 3 and 5 show the accredited system reproducing the reservoir's LOLE with less unserved energy than the reservoir carries. Because unserved energy falls as firm capacity rises, a credit matched on EENS must be smaller than the LOLE-matched credit, and the accredited system then shows more loss-of-load days than the reservoir does: frequent shallow shortfalls in place of rare deep ones. Any scalar credit matches one functional of the shortfall distribution and misstates the others, in a direction set by the difference in depth between a resource that fails at the margin and one that fails whole.

Which functional to match depends on how interruption cost scales. LOLE and EENS correspond to two extreme shapes of that cost: a fixed cost for each day with any interruption, whatever its depth, and a cost proportional to the energy unserved. Customer damage functions lie between, rising with duration and depth at rates that differ by customer class. A reservoir's shortfalls are deep where a firm resource's are marginal, so whatever weight the cost function places on depth, a credit matched on days assigns that weight zero. The factorization of Section 2.4 into frequency, duration, and depth covers the size, frequency, and duration of shortfalls that the Energy Systems Integration Group recommends quantifying {[}6{]}; the identity adds the assignment of each factor to a term.

What follows is not that accreditation should be abandoned, which is not on offer, but that it should be accompanied. The split of {[}1, Sec. 8.3{]} is computable inside any assessment already running a sequential simulation, at no additional sampling cost, and reporting the boundary term's share alongside the credit tells the reader how much of the system's risk the credit failed to represent. Where the share is near zero the credit can be relied on, and the 16-hour row of Table 3 is that case. Where the share is near one the credit is measuring the smaller term, and the procurement decision it informs is the wrong one: the system needs megawatt-hours and is being quoted the price of megawatts. Reporting the share costs nothing and converts a scalar that cannot be checked into one that can.

\subsection*{3.2 A credit set by formula rather than by model}

Section 3.1 treats accreditation as the large systems compute it: a credit derived from a reliability model, published as a duration-indexed table. PJM's marginal ELCC class ratings, the New York Independent System Operator's capacity accreditation factors, and the Midcontinent Independent System Operator's direct loss-of-load method are all of this kind, and Table 3 reproduces the procedure. California does not. It sets storage qualifying capacity by formula, as the maximum continuous energy divided by 4, capped at the power rating, with no reliability model behind the number. Table 6 applies that rule to the system of Section 2.1 and separates two errors that Table 3 reports as one.

\begin{table*}[!t]
\caption{A credit set by formula against a credit set by reliability model, on the scenario of Section 2.1. The formula is the California qualifying-capacity rule: continuous energy divided by 4, capped at the power rating. EENS in GWh/yr. The last two columns are the factor by which each accredited model understates the reservoir's true unserved energy.}
\centering\footnotesize
\begin{tabular}{@{} >{\raggedright\arraybackslash}p{\dimexpr 0.1019\linewidth-2\tabcolsep\relax} >{\raggedright\arraybackslash}p{\dimexpr 0.1019\linewidth-2\tabcolsep\relax} >{\raggedright\arraybackslash}p{\dimexpr 0.0926\linewidth-2\tabcolsep\relax} >{\raggedright\arraybackslash}p{\dimexpr 0.1296\linewidth-2\tabcolsep\relax} >{\raggedright\arraybackslash}p{\dimexpr 0.1112\linewidth-2\tabcolsep\relax} >{\raggedright\arraybackslash}p{\dimexpr 0.1019\linewidth-2\tabcolsep\relax} >{\raggedright\arraybackslash}p{\dimexpr 0.1019\linewidth-2\tabcolsep\relax} >{\raggedright\arraybackslash}p{\dimexpr 0.1296\linewidth-2\tabcolsep\relax} >{\raggedright\arraybackslash}p{\dimexpr 0.1296\linewidth-2\tabcolsep\relax}@{}}
\toprule
Storage duration & Formula credit & Model credit & Over-credit & EENS true & EENS formula & EENS model & Formula understates & Model understates \\
\midrule
\bottomrule
2 h & 6.00 GW (50\%) & 6.06 GW (50\%) & −1\% & 55.71 & 18.55 & 17.85 & 3.0× & 3.1× \\
\textbf{4 h} & \textbf{12.00 GW (100\%)} & \textbf{8.75 GW (73\%)} & \textbf{+37\%} & \textbf{12.99} & \textbf{0.21} & \textbf{2.97} & \textbf{63×} & \textbf{4.4×} \\
8 h & 12.00 GW (100\%) & 11.66 GW (97\%) & +3\% & 0.87 & 0.21 & 0.28 & 4.2× & 3.1× \\
16 h & 12.00 GW (100\%) & 12.01 GW (100\%) & 0\% & 0.21 & 0.21 & 0.20 & 1.0× & 1.0× \\
\end{tabular}
\end{table*}
The two credits agree at 2, 8, and 16 hours and diverge at 4, where the formula awards the full power rating against the model's 73\%. The reason is structural rather than particular to this system. The formula is linear in duration to a kink at 4 hours and flat thereafter, while the accreditation curve is concave and does not saturate until 16 hours; the kink is placed at the duration where the true curve is steepest. At less than 2 hours the formula under-credits instead, by 28\% at 1 hour, because it runs through the origin while the reliability-derived credit does not.

What the divergence costs is the next-to-last column. At 4 hours the formula credits the reservoir at its full power rating in every hour, which is the firm-power representation of Section 2.4---the one that section calls a bound rather than a practice, on the grounds that no planner firm-credits a battery for 8,760 hours. In this scenario, California's tariff does. The accredited model then understates unserved energy by 63 times rather than the 4.4 times a reliability-derived credit produces, and the two errors compound: a credit 37\% too generous, carried into a metric that cannot see the depth of the boundary term at all.

Two qualifications bound the comparison, and both narrow it. The California framework does not end at the qualifying-capacity number: the same rules require an hourly showing across all 24 hours and a charging-sufficiency test, and duration re-enters through both. Neither is modeled here, so the 63-fold figure is what a formula-derived credit alone would deliver, not an estimate of error in that state's resource-adequacy program. And the scenario is Spanish data under a California rule, which is a counterfactual rather than a description of either system. What survives both qualifications is the structural point: where a credit is set by formula, the duration at which the formula bends need not be the duration at which reliability bends, and the decomposition says how much that misalignment costs.

The same diagnosis applies a third time. In {[}1, Sec. 3{]} an unconditional marginal collapsed a joint law and discarded the tail dependence that binds. In {[}1, Sec. 9{]} a scalar value of lost load collapsed a distribution of valuations and discarded which load is curtailed. Here a scalar credit collapses a boundary functional and discards whether the shortfall is deep. Each time the operator was innocent and the argument was not; each time the discarded structure was the structure that governs the shortfall region. The convolution has survived every one of these. The reported number has not.

\section*{4. Conclusion}

On 4 years of Spanish system data, the empty-boundary term supplies the great majority of the risk at the storage durations now being built---98\% of EENS at 4 hours---and vanishes only when the reservoir's energy rating covers the cumulative drawdown of the multi-day droughts, as the drawdown criterion of Part I predicts. The consequence for accreditation is not a matter of degree. ELCC holds fixed a count of days, and a count of days registers how often the binding term produces a shortfall but not how deep it runs, so the credit misstates the binding term by construction; a 4-hour battery meets the one-day-in-ten-years standard while carrying nearly 10 times the unserved energy its accredited model reports, and a credit set by formula rather than by model can compound the error further. The remedy costs nothing: the power-limited term is a quadrature, and reporting the boundary term's share alongside the credit tells a planner how much of the system's risk the credit has failed to represent. Where the share is near zero the credit can be relied on; where it is near one the system needs megawatt-hours and is being quoted the price of megawatts.

\section*{Acknowledgment}

The author used a large language model (Anthropic Claude) to assist with reformatting the manuscript to IEEE style, with language editing, and with drafting connective and framing text. All technical results, derivations, and conclusions are the author's own and were verified by the author.

\section*{Declaration of competing interest}

The author is a Principal Emeritus of The Brattle Group, an economic consultancy whose clients include regulated utilities, regulators, and other participants in electricity markets. The views expressed are the author's own and do not represent the positions of The Brattle Group or of any client. The author received no specific funding for this work and declares no further competing interests.

\section*{Data availability}

The numerical results of Sections 2 and 3 are computed from 4 years of openly published ENTSO-E hourly system data for Spain covering 2015 through 2018---actual hourly load, onshore wind and solar generation, 35,023 complete hours---retrieved via the public unit8co/darts mirror and verified by MD5 checksum. Renewable capacity is scaled by a common factor while the real hourly shapes, and hence the weather-driven joint tails and the multi-day droughts, are preserved. No proprietary or confidential data are used.

\section*{Code availability}

The code reproducing every numerical result and every figure in Sections 2 and 3, and the closed-form benchmark of {[}1, Appendix A{]}, accompanies this paper as supplementary material. It downloads the ENTSO-E record, verifies its checksum before use, and regenerates the decomposition, the accreditation comparison, its control, the calibration to the one-day-in-ten-years standard, and the closed-form comparative statics of {[}1, Appendix A.2{]}. It depends only on NumPy and pandas, exposes its random seed, and is deterministic given it. Results are written both to the console and to machine-readable files.

\end{document}